%% file: main.tex
\documentclass[letterpaper, 10 pt, conference]{Misc/ieeeconf}
\IEEEoverridecommandlockouts                            
\include{Misc/a_macros}

\title{\LARGE \bf
Backup Plan Constrained Model Predictive Control}
\author{Hunmin Kim$^{\dagger}$, Hyungjin Yoon$^{*}$, Wenbin Wan$^{\dagger}$, Naira Hovakimyan$^{\dagger}$, Lui Sha$^{\ddagger}$, and Petros Voulgaris$^{*}$
\thanks{This work has been supported by the National Science Foundation (CNS-1932529) and UIUC STII-21-06.}
\thanks{$^{\dagger}$Hunmin Kim, Wenbin Wan, and Naira Hovakimyan are with the Department of Mechanical Science and Engineering, University of Illinois at Urbana-Champaign, USA.
{\tt\small  \{hunmin, wenbinw2, nhovakim\}@illinois.edu}}%
\thanks{$^{\ddagger}$Lui Sha is with the Department of Computer Science, University of Illinois at Urbana-Champaign, USA.
{\tt\small  lrs@illinois.edu }}%
\thanks{$^{*}$Hyungjin Yoon and Petros Voulgaris are with the Department of Mechanical Engineering, University of Nevada, Reno, USA.
{\tt\small  \{hyungjiny, pvoulgaris\}@unr.edu }}
}

\begin{document}
\maketitle
\thispagestyle{empty}
\pagestyle{empty}
%%%%%%%%%%%%%%%%%%%%%%%%%%%%%%%%%%%%%%%%%%%%%%%%%%%%%%%%%%%%%%%%%%%%%%%%%%%%%%%%

%%%%%%%%%%%%%%%%%%%%%%%%%%%%%%%%%%%%%%%%%%%%%%%%%%%%%%%%%%%
%%%%%%%%%%%%%%%%%%    Abstract       %%%%%%%%%%%%%%%%%%%%%%
%%%%%%%%%%%%%%%%%%%%%%%%%%%%%%%%%%%%%%%%%%%%%%%%%%%%%%%%%%%

\begin{abstract}
This article proposes a new safety concept: backup plan safety. The backup plan safety is defined as the ability to complete one of the alternative missions in the case of primary mission abortion. To incorporate this new safety concept in control problems, we formulate a feasibility maximization problem that adopts additional (virtual) input horizons toward the alternative missions on top of the input horizon toward the primary mission. Cost functions for the primary and alternative missions construct multiple objectives, and multi-horizon inputs evaluate them. To address the feasibility maximization problem, we develop a multi-horizon multi-objective model predictive path integral control (3M) algorithm. Model predictive path integral control (MPPI) is a sampling-based scheme that can help the proposed algorithm deal with nonlinear dynamic systems and achieve computational efficiency by parallel computation. Simulations of the aerial vehicle and ground vehicle control problems demonstrate the new concept of backup plan safety and the performance of the proposed algorithm.
\end{abstract}

\maketitle

\section{Motivation and Introduction}

Traditional path planning problems for robotics and autonomous ground/aerial vehicles consider collision avoidance enough for safety. However, this consideration is not enough for emerging automated systems that perform complex tasks requiring safety criticality if we recall the incident of Miracle on the Hudson (US Airways Flight 1549)~\cite{marra2009migratory}. After a bird strike resulted in all engines' failure, Captain Sullenberger flew along the Hudson river, checking the feasibility of safer landing points. Additionally, the airplane ditched near boats, and this expedited rescue. This observation calls for a need for new safety definitions to cope with mission uncertainties. This paper aims to examine a novel control algorithm considering a new safety definition, which we call {\em  backup plan safety}. The backup plan safety is defined as the ability to complete one of the alternative missions in the case of primary mission abortion. This safety definition will be particularly useful for automated systems that require a long horizon emergency response, such as aerial and nautical transportation systems; and for systems operating under mission uncertainties such as robotics, manufacturing systems, and autonomous vehicles.

\begin{figure}[t]
    \centering
    \includegraphics[width=.45\textwidth]{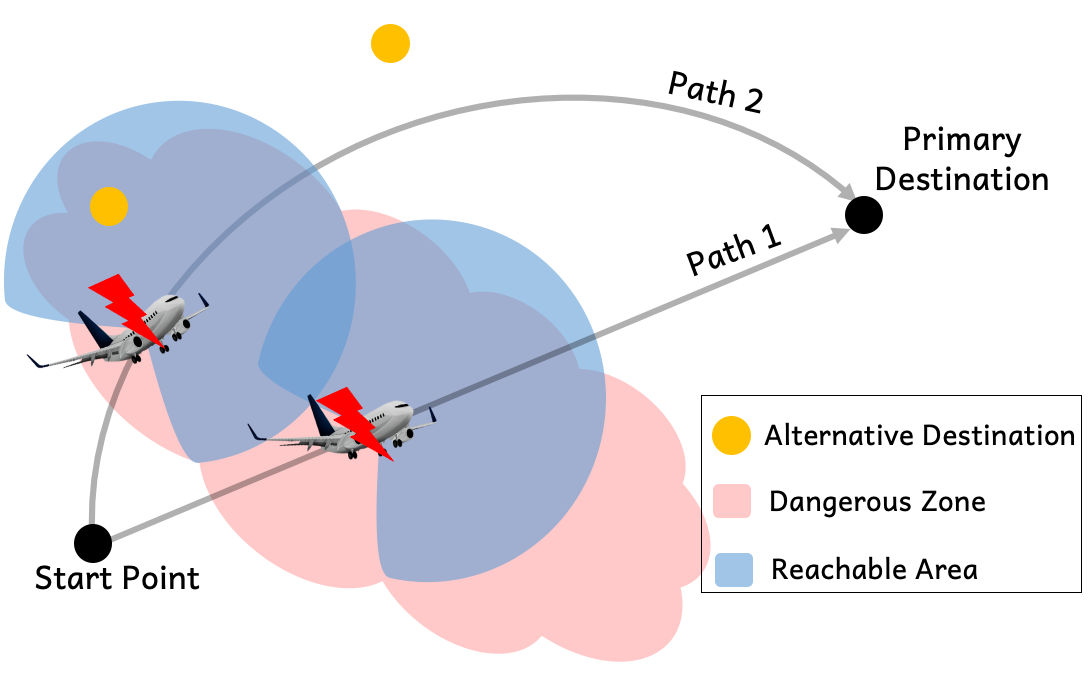}
    \caption{(Feasibility maximization scenario) An aircraft is passing through a dangerous zone (red). The blue area indicates the reachable area after the airplane is under emergency.}
    \label{fig:demo}\vspace{-0.8cm}
\end{figure}

A similar safety concept had been used in aircraft path planning. The Federal Aviation Administration set up the 60-minute rule in 1953, which allows twin-engine aircraft to fly routes no further than 60 minutes from the nearest airport suitable for an emergency landing with the aircraft's speed with one engine being inoperative. To fly outside of the 60-minute distance, one needs to follow extended-range twin-engine operational performance standards (ETOPS). Since then, the 60-min rule with ETOPS has been evaluated to enhance the safety of aircraft~\cite{desantis2013engines}. In particular, consider a scenario that an airplane passes through a dangerous area (e.g., a storm or a bird habitat), shown in Figure~\ref{fig:demo}, toward its destination. A typical path planning solution would be finding the shortest path (Path 1). However, following this path, the airplane has no viable means to safely land at an airport when it confronts an emergency, possibly with limited performance. Path 2 tries to maximize the feasibility of the safe landing given two alternative destinations. Even when an emergency takes place, one of the two alternative destinations is still feasible for the airplane. The backup plan safety proposed in the current paper is a generalization of such rule in terms of the safety standard and application domains.

Model predictive control (MPC) is a general iterative optimal control methodology for a finite control horizon, satisfying a set of constraints. MPC has shown its superiority in the path and motion planning domain for stability and safety~\cite{paden2016survey,aggarwal2020path}. MPC can be implemented for real-time use due to the advancement of computing hardware and algorithmic developments. Recent years have seen efforts regarding integrating machine learning with MPC methods towards establishing a unified framework for learning-based planning and control for enhanced safe autonomy~\cite{aswani2013provably,hewing2019cautious}. Despite those algorithmic developments, the safety destination in literature is yet limited to collision avoidance. The development of multi-objective MPC (MMPC)~\cite{bemporad2009multiobjective} could provide a basis for multi-mission control problems for backup plan safety, where each cost function is associated with a mission. MMPC shows its effectiveness in various applications such as power converter control~\cite{hu2013multi,hu2018multi}, HVAC~\cite{ascione2017new}, and cruise control~\cite{zhao2017real,li2010model}. However, a single prediction horizon used in MPC and MMPC is not enough to address backup plan safety, because cost functions for alternative missions are well evaluated only when the trajectories toward the corresponding mission are given (e.g., mission feasibility).

Stochastic model predictive control (SMPC) exploits the probabilistic uncertainty model in an optimal control problem formulation. SMPC balances the tradeoff between optimizing control objectives and satisfying chance constraints. Typical strategies to address SMPC include stochastic-tube~\cite{cannon2010stochastic,cannon2012stochastic}, stochastic programming~\cite{blackmore2010probabilistic}, and sampling-based approaches~\cite{visintini2006monte,kantas2009sequential,williams2016aggressive,williams2017model, williams2018information}. The current paper focuses on a sampling-based approach to handle nonlinear dynamic systems and enable efficient parallel computing using Graphics
Processing Units (GPUs). In particular, our algorithm is based on model predictive path integral control (MPPI)~\cite{williams2016aggressive,williams2017model, williams2018information} that relies on a generalized importance sampling scheme.

\textbf{Contribution.} The current article proposes a new safety concept - backup plan safety, to enhance the operation of complex autonomous systems under mission uncertainties. The backup plan constrained control problem is formulated as a feasibility maximization problem: MMPC with multi-horizon inputs. We develop the multi-objective multi-horizon model predictive path integral control (3M) algorithm to address the feasibility maximization problem. In particular, MMPC with multi-horizon inputs can handle the multi-mission problem, in which multi-horizon inputs help evaluate all cost functions. By enabling parallel computation using GPUs, MPPI control expedites the computing speed of complex optimization problems with multi-horizon inputs. MPPI control further helps to avoid harmful computation delays by finishing computation in a designated time. Simulations of aerial vehicle and ground vehicle control problems present the new safety concept and the performance of the proposed algorithm.

The remainder of the paper is organized as follows. Section~\ref{sec:MMPC} formulates the feasibility maximization problem with multi-horizon inputs toward the primary and alternative missions. Section~\ref{sec:3M} introduces MPPI control and proposes the 3M algorithm to address the feasibility maximization problem. Simulation results of an aerial vehicle and a ground vehicle are presented in Section~\ref{sec:sim}.

\section{Feasibility Maximization}\label{sec:MMPC}
Consider the discrete-time switched dynamic systems:
\begin{align}
    x_{k+1}= f(x_k,u_k,j),
    \label{eq:sysmodel}
\end{align}
where $f:\RR^{n_x}\times\RR^{n_u}\times \mathbb{N}\rightarrow \RR^{n_x}$ is a nonlinear system function, $x_k \in \RR^{n_x}$ is the system state, $u_k \in \RR^{n_u}$ is the control input at time $k \geq 0$.
Mode index $j \in \mathcal{J} \subset \mathbb{N}$ determines the function $f$, where $\mathcal{J}$ is the mode index set.

\begin{remark}
{The switched system model in~\eqref{eq:sysmodel} can represent changing dynamic models depending on time and events during the operation. For instance, in the aircraft control problem in Figure~\ref{fig:demo}, the dynamic system model moving toward alternative destinations can be a single-engine failure model.}\oprocend
\end{remark}

We assume that the control authority knows one primary mission and $m$ alternative missions. It is said that the mission $i$ is completed at time $t$ if
\begin{align}
   t=&\argmin_k k\nnum\\
   &{\rm \ s.t. \ } d(x_k,p^i)=0 ,
   \label{eq:MissionCompletetion}
\end{align}
where $d:\RR^{n_x}\times\RR^{n_x}\rightarrow \RR_{\geq0}$ is the distance metric between the internal system state $x_k$ and mission state $p^i \in \RR^{n_x}$. The control objective for the system~\eqref{eq:sysmodel} is to complete the primary mission, and if the primary mission is aborted in the middle of the operation, then the system should complete one of the alternative missions instead. %We call it backup plan safety.

An alternative mission may not align with the primary mission. Therefore, the control problem should balance between the primary and alternative cost functions.
Considering this fact, we formulate the control problem of the system~\eqref{eq:sysmodel} as a feasibility maximization:
\begin{align}
    &\min_{\mathbf{U}} \mathbf{J}(\mathbf{X},\mathbf{U})\nnum\\
    &{s. t. \ } x_{k+1}= f(x_k,u_k,j),
    \label{eq:mmpc0}
\end{align}
where $\mathbf{X} \in \RR^{(N+1+\frac{N(N-1)m}{2})n_x}$ is the collection of states, $\mathbf{U} \in \RR^{(N+\frac{N(N-1)m}{2})n_u}$ is the collection of inputs, and $\mathbf{J}: \RR^{(N+1+\frac{N(N-1)m}{2})n_x} \times \RR^{(N+\frac{N(N-1)m}{2})n_u} \rightarrow \RR^{m+1}$ is an $m+1$ dimensional vector cost function. Each cost function is associated with a primary or alternative mission. The problem contains finite $N \in {\mathbb{N}}$ prediction horizon control inputs at each time step $t$ (i.e., $[t,t+N-1]$) that minimize the pre-designed multiple cost functions. After executing the first control input, the prediction horizon will be shifted forward and optimize the cost function again as in the standard MPC. We assume that the cost function $\mathbf{J}$ includes soft constraints. In particular, one could use the Lagrange multiplier method to realize state/input constraints and feasibility guarantee.

The key difference of the feasibility maximization problem~\eqref{eq:mmpc0} from the multi-objective MPC in~\cite{bemporad2009multiobjective} is that the input $\mathbf{U}$ in~\eqref{eq:mmpc0} includes an input horizon toward the primary mission and additional (virtual) input horizons toward alternative missions. The multi-horizon control inputs help  to evaluate the cost function toward both the primary and alternative missions. Section~\ref{sec:input} discusses the multi-horizon control inputs $\mathbf{U}$, state trajectories $\mathbf{X}$, and their dimensions. Section~\ref{sec:cost} discusses the cost function $\mathbf{J}$.

The feasibility maximization problem in~\eqref{eq:mmpc0} is a family of multi-objective optimization problems, where the elements of cost functions $\mathbf{J}(\mathbf{X},\mathbf{U})$ are conflicting with each other in general, and there is no solution that minimizes all the cost functions simultaneously. As in multi-objective optimization~\cite{marler2004survey,bemporad2009multiobjective}, Pareto optimality plays a key role in defining optimality.

\begin{definition}[Pareto optimality\cite{bemporad2009multiobjective}]
A feasible solution is Pareto optimal if and only if it is not dominated by any other feasible solution. In other words, feasible solution $\mathbf{V}$ is Pareto optimal if and only if there is no feasible $\mathbf{U}$ such that $\mathbf{J}(\mathbf{X},\mathbf{U}) \leq \mathbf{J}(\mathbf{X},\mathbf{V})$ holds element-wise, and strict inequality holds for at least one element.
\end{definition}

Multi-objective genetic algorithm~\cite{konak2006multi} is particularly useful to identify Pareto optimal sets and corresponding Pareto frontier, the set of Pareto optimal cost function values. However, they may not be used in real-time applications due to the computational complexity. In this case, we can assign parameterized weights to the cost functions so that we limit our focus to the particular subset of solutions. With the weight assignment, the feasibility maximization problem~\eqref{eq:mmpc0} can be reformulated by
\begin{align}
    &\min_{\mathbf{U}} \alpha^\top\mathbf{J}(\mathbf{X},\mathbf{U})\nnum\\
    &{s. t. \ } x_{k+1}= f(x_k,u_k,j),
    \label{eq:mmpc}
\end{align}
where the weight vector $\alpha = [\alpha^0,\cdots,\alpha^m]^\top \in \RR^{m+1}$
satisfies $\alpha^i \in [0,1] \subset \RR$ and $\sum_{i=0}^{m}\alpha^i=1$.
% without loss of generality. 
The solution of~\eqref{eq:mmpc} is a Pareto optimal solution of~\eqref{eq:mmpc0} if $\alpha^i>0$ for $\forall i$~\cite{boyd2004convex}, while this may not hold when there exist some indices such that $\alpha^i=0$. The choice of $\alpha$ governs the valuation on each mission, resulting in different optimal control sequences. Section~\ref{sec:weight} discusses how to choose the weight vector $\alpha$.

\subsection{Multi-horizon Inputs and Trajectories}\label{sec:input}

The control input
\begin{align}
  \mathbf{U} = [(\mathbf{U}^0)^\top,(\mathbf{U}^1)^\top,\cdots,(\mathbf{U}^m)^\top]^\top
  \label{eq:BigU}
\end{align}
consists of inputs toward the primary mission
\begin{align*}
    \mathbf{U}^0 = [u_0^\top,u_1^\top,\cdots,u_{N-2}^\top,u_{N-1}^\top]^\top
\end{align*}
and additional (virtual) input horizons toward the alternative missions $\mathbf{U}^i$ for $i=1,\cdots,m$, where $\mathbf{U}^0$ is the input horizon used in the standard MPC.

Two important properties should be considered when defining the inputs toward the alternative missions $\mathbf{U}^i$:
\begin{itemize}
    \item It is unknown when the system will abort the mission;
    \item The control horizon toward the alternative missions should be $N$.
\end{itemize}
It is required for $\mathbf{U}^i$ to consider the possibility of mission abortion at every point for the first property. Accordingly, We construct $\mathbf{U}^i$ as
\begin{align}
    \mathbf{U}^i = [(\mathbf{U}_0^i)^\top,(\mathbf{U}_1^i)^\top,\cdots,(\mathbf{U}_{N-3}^i)^\top,(\mathbf{U}_{N-2}^i)^\top]^\top,
    \label{eq:MediumU}
\end{align}
where $\mathbf{U}_p^i$ is the control sequence toward the $i^{th}$ alternative mission when the primary mission is aborted after $u_p$ has been executed. By this definition, the first $p+1$ control inputs of $\mathbf{U}_p^i$ are $u_0,\cdots,u_p$. Because the dimension of $\mathbf{U}_p^i$ should be $\mathbf{U}_p^i \in \RR^{N\times n_u}$ to satisfy the second property, there are additional $N-(p+1)$ number of inputs toward the alternative mission $i$, in $\mathbf{U}_p^i$.
Then, we have
\begin{align*}
    \mathbf{U}_p^i = [u_0^\top,\cdots,u_p^\top,(u_{p,p+1}^i)^\top,\cdots,(u_{p,N-1}^i)^\top]^\top,
\end{align*}
where $u_{p,q}^i$ is the $q^{th}$ control input when we decide to abort the primary mission after executing $u_p$ and choose $i^{th}$ alternative mission for the next.

\begin{figure}[htp]
\begin{subfigure}{\linewidth}
  \centering
  \includegraphics[width=\linewidth]{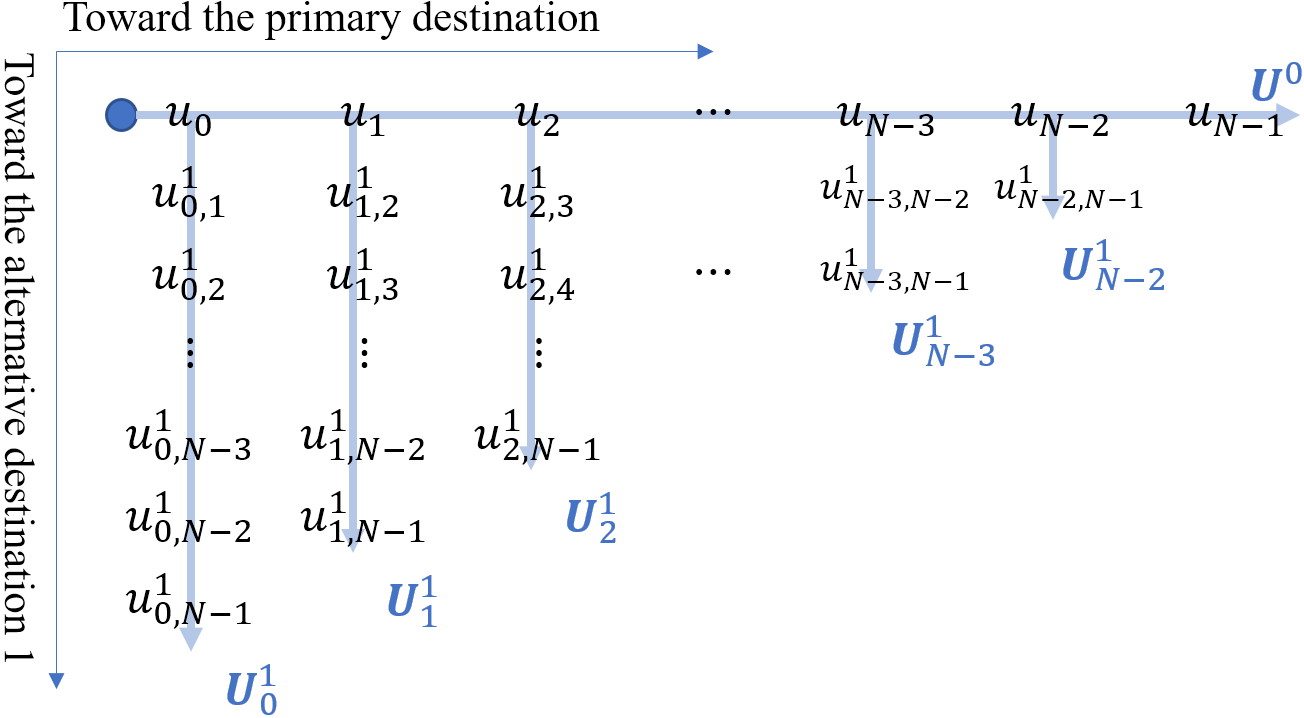}
  \caption{The elements of input $\mathbf{U}$ are presented in an array. To construct $\mathbf{U}^0$ or $\mathbf{U}_p^1$, one can start at the blue dot at the left upper corner, and move right until the primary mission is aborted, and move downside after the primary mission is aborted and the alternative mission $1$ is chosen for the next mission.}
\end{subfigure}
\begin{subfigure}{\linewidth}
  \centering
  \includegraphics[width=\linewidth]{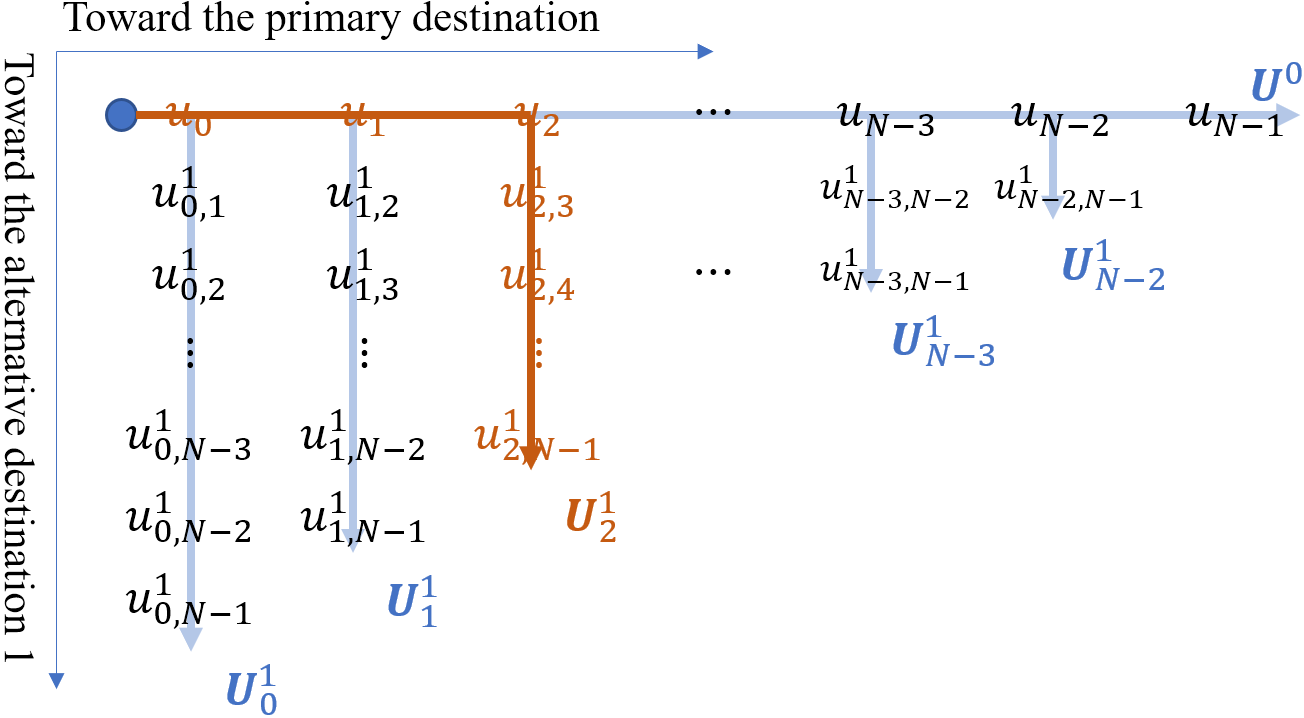}
  \caption{An example of the way to construct $\mathbf{U}_2^1=[u_0^\top,u_1^\top,u_2^\top,(u_{2,3}^1)^\top,(u_{2,4}^1)^\top,\cdots,(u_{2,N-1}^1)^\top{]}^\top$ (marked in red). The first three elements are the elements of $\mathbf{U}^0$.}
\end{subfigure}
\caption{Visualization of inputs $\mathbf{U}$ when $m=1$.}
\label{fig:inputs}
\end{figure}

Input $\mathbf{U}$ is visualized in Figure~\ref{fig:inputs} when $m=1$, which presents the relation between $\mathbf{U}$, $\mathbf{U}^0$, and $\mathbf{U}_p^1$, and shows how $u_p$, and $u_{p,q}^i$ can construct $\mathbf{U}^0$ and $\mathbf{U}_p^1$.

The states $\mathbf{X}= [(\mathbf{X}^0)^\top,(\mathbf{X}^1)^\top,\cdots,(\mathbf{X}^m)^\top]^\top$ are constructed by simulating the input $\mathbf{U}$ on the system~\eqref{eq:sysmodel}, where $\mathbf{X}^0$ and $\mathbf{X}^i=[(\mathbf{X}_0^i)^\top,\cdots,(\mathbf{X}_{N-2}^i)^\top]^\top$ are the simulated states of inputs $\mathbf{U}^0$ and $\mathbf{U}^i$, respectively for $i=1,\cdots,m$.
Likewise, $\mathbf{X}_p^i = [x_0^\top,\cdots,x_p^\top,(x_{p,p+1}^i)^\top,\cdots,(x_{p,N}^i)^\top]^\top$ is the simulated state of the input $\mathbf{U}_p^i$, for $i=1,\cdots,m$ and $p=0,\cdots,N-2$.

\begin{remark}[Dimension of states and inputs]
It is worth re-emphasizing that the first $p+1$ elements of $\mathbf{U}_p^i$ are those of $\mathbf{U}^0$. Correspondingly, the first $p+2$ elements of $\mathbf{X}_p^i$ are $x_0,x_1,\cdots,x_p,x_{p+1}$, which are the elements of $\mathbf{X}^0$. Therefore, the input $\mathbf{U}$ consists of the $N$ number of inputs toward the primary mission in $\mathbf{U}^0$, and $N-1,N-2,\cdots,1$ additional number of inputs toward the alternative mission $i$ in $\mathbf{U}_0^i,\mathbf{U}_1^i,\cdots,\mathbf{U}_{N-2}^i$. Therefore, $\mathbf{U}$ consists of $N+\frac{N(N-1)}{2}m$ independent elements, and $\mathbf{X}$ consists of $N+1+\frac{N(N-1)}{2}m$ independent elements. The dimensions of independent input and state variables are large compared to the standard MPC. We will deal with the induced computational complexity by using MPPI control described in Section~\ref{sec:MPPI}.
\oprocend\label{rmk:1}
\end{remark}

\subsection{Multi-objective Cost Functions}\label{sec:cost}
The cost function 
\begin{align*}
    \mathbf{J}(\mathbf{X},\mathbf{U}) \triangleq [\mathbf{J}^0(\mathbf{X}^0,\mathbf{U}^0),\mathbf{J}^1(\mathbf{X}^1,\mathbf{U}^1),\cdots,\mathbf{J}^m(\mathbf{X}^m,\mathbf{U}^m)]^\top
\end{align*}
is an $m+1$ dimensional vector function, where the elements are the average cost over state-input trajectories toward the corresponding mission:
\begin{align*}
    &\mathbf{J}^0(\mathbf{X}^0,\mathbf{U}^0) =J^0(\mathbf{X}^0,\mathbf{U}^0)\nnum\\
    &\mathbf{J}^i(\mathbf{X}^i,\mathbf{U}^i) =
    \frac{1}{N-1}\sum_{p=0}^{N-2}J^i(\mathbf{X}_p^i,\mathbf{U}_p^i)
\end{align*}
for $i=1,\cdots,m$. The function $J^i$ is a standard cost function for the mission $i$:
\begin{align}
    J^i(\mathbf{X}^i,\mathbf{U}^i) = \sum_{k=1}^{N} L^i(\mathbf{X}^i(k),\mathbf{U}^i(k)) + F^i(\mathbf{X}^i(N+1))
    \label{eq:costSmallJ}
\end{align}
that consists of the cost-to-go $L^i$ and the terminal cost $F^i$, where $\mathbf{X}^i(k) \in \RR^{n_x}$ and $\mathbf{U}^i(k) \in \RR^{n_u}$ are the $k^{th}$ state and input of $\mathbf{X}^i$ and $\mathbf{U}^i$, respectively.

\subsection{Weight Vector}\label{sec:weight}
%{\color{red}(This is not a sufficient condition, but necessary condition. Another necessary condition is to guarantee $x=p^0$ is the unique solution of $\alpha^\top (x) \mathbf{J}(x,0)$; we may need some revision for this paragraph)}

When choosing the weight vector $\alpha$ in the feasibility maximization~\eqref{eq:mmpc}, we should consider two important issues. First of all, for any fixed $\alpha$ (except for $\alpha=[1,0,\cdots,0]^\top$), there may not exist Pareto optimal that achieves the primary mission because the feasibility maximization problem~\eqref{eq:mmpc} tries to minimize the weighted sum of the cost functions of the primary and alternative missions. Second, the choice of $\alpha$ will affect the closed-loop stability and performance. In what follows, we design the desired weight vector $\alpha_d$ to address the first issue such that it converges to $\alpha=[1,0,\cdots,0]^\top$ as the system is about to achieve the primary mission. Furthermore, we will adopt the optimization-based weight update law for closed-loop stability.

The desired weight vector $\alpha_d(x_t)$ is designed as follows:
% \begin{align}     \alpha_d(x_t) &= [1-\gamma+\frac{\gamma/d(x_t,p^0)}{\sum_{i=0}^m(\frac{1}{d(x_t,p^i)})}, \frac{\gamma/d(x_t,p^1)}{\sum_{i=0}^m(\frac{1}{d(x_t,p^i)})},\nnum\\     &\cdots,\frac{\gamma/d(x_t,p^m)}{\sum_{i=0}^m(\frac{1}{d(x_t,p^i)})}]^\top     \label{eq:alphad} \end{align}
\begin{align}
    \alpha_d(x_t) &= [1-\gamma+w^0(x_t) \gamma,w^1(x_t) \gamma,\cdots, w^m(x_t) \gamma]^\top,
    \label{eq:alphad}
\end{align}
where
\begin{align*}
    w^i=\frac{exp(-\frac{1}{\lambda_{\alpha} }d(x_t,p^i))}{\sum_{l=0}^m exp(-\frac{1}{\lambda_{\alpha}}d(x_t,p^l))}
\end{align*}
is the Gibbs distribution with temperature parameter $\lambda_{\alpha}$, and $0 < 1-\gamma \leq 1$ is the pre-determined weight on the primary mission. Function $d$ is the distance metric between the state and mission state. It can be verified that $\alpha_d$ in~\eqref{eq:alphad} satisfies the constraints on $\alpha$ (i.e., $\alpha_d^\top \mathbf{1}=1$ and $\alpha_d \in [0,1] \subset \RR$), and converges to $[1,0,\cdots,0]^\top$ as $d(x_t,p_0)$ decreases to zero. 

Let us define $\alpha_t$ and $\alpha_{t-1}$ as $\alpha$ in~\eqref{eq:mmpc} chosen at the current time $t$ and the one at the previous time $t-1$, respectively. Now, $\alpha_t$ must be chosen close to its desired value $\alpha_d(x_t)$, while its choice guarantees closed-loop stability as in~\cite{bemporad2009multiobjective}:
\begin{align}
    \alpha_t = &\argmin_{\alpha}h(\alpha-\alpha_d(x_t))\nnum\\
    &{\rm s.t. \ } \alpha^\top (\mathbf{J}(\hat{\mathbf{X}}_{t-1},\hat{\mathbf{U}}_{t-1})-\mathbf{F}(\hat{\mathbf{X}}_{t-1},\hat{\mathbf{U}}_{t-1})) \nnum\\
    &\quad \leq \alpha_{t-1}^\top (\mathbf{J}(\hat{\mathbf{X}}_{t-1},\hat{\mathbf{U}}_{t-1})-\mathbf{F}(\hat{\mathbf{X}}_{t-1},\hat{\mathbf{U}}_{t-1})),\nnum\\
    &\quad \alpha^\top\mathbf{1}=1,\nnum\\
    &\quad \alpha_i \geq 0, \ i = 0,\cdots,m ,
    \label{eq:alphaup}
\end{align}
where $h$ is a convex cost function that penalizes the difference between $\alpha_t$ and $\alpha_d(x_t)$.

As in~\eqref{eq:BigU} and~\eqref{eq:MediumU}, the input $\hat{\mathbf{U}}_{t-1}$ is defined by $\hat{\mathbf{U}}_{t-1}=[(\hat{\mathbf{U}}_{t-1}^0)^\top,(\hat{\mathbf{U}}_{t-1}^1)^\top,\cdots,(\hat{\mathbf{U}}_{t-1}^m)^\top]^\top$, where $\hat{\mathbf{U}}_{t-1}^i = [(\hat{\mathbf{U}}_{0,t-1}^i)^\top,\cdots,(\hat{\mathbf{U}}_{N-2,t-1}^i)^\top]^\top$ for $i=1,\cdots,m$. The input $\hat{\mathbf{U}}_{t-1}$ is constructed from the optimal control input at the previous step $\mathbf{U}_{t-1}^i$. In particular, we have $\hat{\mathbf{U}}_{t-1}^0$ and $\hat{\mathbf{U}}_{p,t-1}^i$ by removing the first input $u_0$ (which has been already executed) and appending a zero vector $\mathbf{0}_{n_u} \in \RR^{n_u}$ at the end to $\mathbf{U}_{t-1}^0$ and $\mathbf{U}_{p,t-1}^i$, i.e.,
\begin{align}
    \hat{\mathbf{U}}_{t-1}^0 &= [u_1^\top, \cdots, u_{N-1}^\top, \mathbf{0}_{n_u}^\top]^\top\nnum\\
    \hat{\mathbf{U}}_{p,t-1}^i &= [u_1^\top, \cdots, u_p^\top, (u_{p,p+1}^i)^\top,\cdots, (u_{p,N-1}^i)^\top, \mathbf{0}_{n_u}^\top]^\top
    \label{eq:uhat}
\end{align}
for $i=1,\cdots,m$ and $p=1,\cdots,N-2$.
The state $\hat{\mathbf{X}}_{t-1}$ is the corresponding simulated trajectory of the input $\hat{\mathbf{U}}_{t-1}^i$. The vector function
\begin{align*}
&\mathbf{F}(\hat{\mathbf{X}}_{t-1},\hat{\mathbf{U}}_{t-1}) \nnum\\
&= [
L^0(\hat{\mathbf{X}}_{t-1}^0(N),\hat{\mathbf{U}}_{t-1}^0(N))+F^0(\hat{\mathbf{X}}_{t-1}^0(N+1)),\nnum\\
&\sum_{p=0}^{N-2}\frac{L^1(\hat{\mathbf{X}}_{p,t-1}^1(N),\hat{\mathbf{U}}_{p,t-1}^1(N))+F^1(\hat{\mathbf{X}}_{p,t-1}^1(N+1))}{N-1},\nnum\\
&\cdots,\nnum\\
&\sum_{p=0}^{N-2}\frac{L^m(\hat{\mathbf{X}}_{p,t-1}^m(N),\hat{\mathbf{U}}_{p,t-1}^m(N))+F^m(\hat{\mathbf{X}}_{p,t-1}^m(N+1))}{N-1}]^\top
\end{align*}
%$\mathbf{F}(\hat{\mathbf{X}}_{t-1},\hat{\mathbf{U}}_{t-1}) = [L^0(\hat{\mathbf{X}}_{t-1}^0(N),\hat{\mathbf{U}}_{t-1}^0(N))+F^0(\hat{\mathbf{X}}_{t-1}^0(N+1)),\frac{1}{N-1}\sum_{p=0}^{N-2}(L^1(\hat{\mathbf{X}}_{p,t-1}^1(N),\hat{\mathbf{U}}_{p,t-1}^1(N))+F^1(\hat{\mathbf{X}}_{p,t-1}^1(N+1))),\cdots,\frac{1}{N-1}\sum_{p=0}^{N-2}(L^m(\hat{\mathbf{X}}_{p,t-1}^m(N),\hat{\mathbf{U}}_{p,t-1}^m(N))+F^m(\hat{\mathbf{X}}_{p,t-1}^m(N+1)))]$
is the last cost-to-go function and terminal cost. Now that the last control input in $\hat{\mathbf{U}}_{t-1}$ is a dummy $\mathbf{0}_{n_u}$, we should not take into account the cost incurred by the dummy. The first constraint in~\eqref{eq:alphaup} implies that $\alpha_t$ should be chosen such that the value function $\alpha^\top (\mathbf{J}(\hat{\mathbf{X}}_{t-1},\hat{\mathbf{U}}_{t-1})-\mathbf{F}(\hat{\mathbf{X}}_{t-1},\hat{\mathbf{U}}_{t-1}))$ is decreasing for the closed-loop stability. If the cost function $h$ is quadratic, then the problem~\eqref{eq:alphaup} becomes a quadratic programming problem because all the constraints are linear with respect to the decision variable $\alpha$.

\section{Multi-objective Multi-horizon Model Predictive Path Integral Control (3M)}\label{sec:3M}

The current section proposes a 3M algorithm to address the feasibility maximization problem~\eqref{eq:mmpc}.
Given that the system model~\eqref{eq:sysmodel} is nonlinear, and the dimension of input $\mathbf{U}$ is large, the proposed algorithm is based on MPPI control that is a sampling-based and parallel computable MPC. Section~\ref{sec:MPPI} introduces MPPI control, and Section~\ref{sec:3M} presents the 3M algorithm.

\subsection{Model Predictive Path Integral Control (MPPI)}\label{sec:MPPI}

The MPPI control algorithm solves stochastic optimal control problems based on the (stochastic) sampling of the system trajectories through parallel computation~\cite{williams2016aggressive,williams2017model, williams2018information}.
Due to the sampling nature, the algorithm does not require derivatives of either the dynamics or the cost function of the system, which enables to handle nonlinear dynamics and non-smooth/non-differentiable cost functions without approximations. With the help of GPUs for expediting the parallel computation, the MPPI can be implemented in real-time even for relatively large dimensions of the state space (e.g., there are 48 state variables for the 3-quadrotor control example in \cite{williams2017model}). The computational efficiency from paralleled stochastic sampling and the ability to directly handle non-smooth cost functions make MPPI appealing for real-time control problems.

Consider the dynamic system~\eqref{eq:sysmodel}
with a fixed $j$ (and thus omitted) and a noise corrupted input:
\begin{align*}
    x_{k+1}= f(x_k,u_k+\epsilon_k),
\end{align*}
where $\epsilon_k \in \RR^{n_u}$ is an independent and identically distributed (i.i.d.) Gaussian noise, i.e., $\epsilon_k \sim \mathcal{N}(0, \Sigma)$ with the known co-variance matrix $\Sigma$.
Given a finite time horizon $N$, the goal of the optimization problem is to find an input trajectory $\mathbf{U}^0= [u_0^\top, u_1^\top, \cdots, u_{N-1}^\top]^\top$ that minimizes the expected cost over all trajectories:
\begin{align*}
        \mathbf{U}^* = \argmin_{\mathbf{U}^0} \mathbb{E} [S(\tau)],
\end{align*}
where $\tau = \{x_0, u_0, x_1, u_1,\cdots, u_{N-1}, x_N\}$. The cost function of a trajectory is given as follows:
\begin{align}\label{eq:costMPPI}
    S(\tau) = \phi(x_N) + \sum_{t=0}^{N-1} \left( c(x_t) + \lambda u_t^\top\Sigma^{-1}\epsilon_t  \right),
\end{align}
where $\phi(x_N)$ is the terminal cost, $c(x_t)$ is the state-dependent running cost, and $\lambda$ is a parameter discussed later. The cost function~\eqref{eq:costMPPI} depends on unknown random variable $\epsilon_k$ for $k=0,\cdots,N-1$. MPPI control relies on a sampling-based method to evaluate the cost~\eqref{eq:costMPPI}. In particular, we sample $\epsilon_k$ from the distribution $\mathcal{N}(0, \Sigma)$, and construct $K$ trajectories of noises $\boldsymbol\epsilon^q = [(\epsilon_0^q)^\top,(\epsilon_1^q)^\top,\cdots,(\epsilon_{N-1}^q)^\top]$ for $q=1,\cdots,K$. The cost function $S$ can be evaluated for each trajectory $\boldsymbol\epsilon^q$.
Furthermore, MPPI control adopts the iterative update law~\cite{williams2017model} to obtain the current optimal input $\mathbf{U}^0$ around the  previous optimal input $\mathbf{U}_{t-1}^0$ as follows:
\begin{align*}
     \mathbf{U}^0 = [u_{1,t-1}^\top,u_{2,t-1}^\top,\cdots,u_{N-1,t-1}^\top,\mathbf{0}^\top]^\top +\sum_{q=1}^Kw^q{\boldsymbol\epsilon}^q,
\end{align*}
where
\begin{align*}
    w^q=\frac{exp(-\frac{1}{\lambda }S(\tau^q))}{\sum_{q=1}^K exp(-\frac{1}{\lambda}S(\tau^q))}
\end{align*}
with $\lambda$ known as the temperature parameter of the Gibbs distribution (or Softmax function), and $\tau^q$ as state-input sets for $q^{th}$ noise trajectory. Input $u_{i,t-1}$ is the $i^{th}$ input of the previous optimal input $\mathbf{U}_{t-1}^0$.

\begin{algorithm}[th]
\caption{3M algorithm}\label{alg:1}
\textbf{Choose tuning parameters:}\\
$K$: Number of sample trajectories;\\
$N$: The size of control horizon;\\
$m$: The number of alternative missions;\\
$\Sigma$: Co-variance of the noise $\epsilon_k$;\\
$\lambda_{\alpha}$, $\lambda$: Temperature parameter of the Gibbs distribution;\\
$\gamma$: Weight on alternative missions;\\
$\mathbf{U}_{0}$: Initial input sequence;
%\algorithmicrequire $K$, $N$. $\lambda$, $\sigma$ \\
%\algorithmicensure 
\begin{algorithmic}[1]
%\For {$k =1$ to $N$}
%=====================================================
%\LineComment{Explanation here}
%=====================================================
\State Measure current state $x_t$ and get $\alpha_d(x_t)$;
\State Obtain $\hat{\mathbf{U}}_{t-1}^0$ and $\hat{\mathbf{U}}_{p,t-1}^i$ in~\eqref{eq:uhat} for $i=1,\cdots,m$ and $p=1,\cdots,N-2$ from $\mathbf{U}_{t-1}$;
\State Update $\alpha_t$ by~\eqref{eq:alphaup};

\State Sample $K$ trajectories of noise $\boldsymbol\epsilon^q$ as in~\eqref{eq:Ksamplenoise};
\State Simulate $\hat{\mathbf{U}}_{t-1}+\boldsymbol\epsilon^q$ on the system~\eqref{eq:sysmodel} to get
$\hat{\mathbf{X}}_{t-1}^q$ for $q=1,\cdots,K$;
\State Evaluate $\alpha_t^\top \mathbf{J}^q$ for $q=1,\cdots,K$, where $\mathbf{J}^q = \mathbf{J}(\hat{\mathbf{X}}_{t-1}^q,\hat{\mathbf{U}}_{t-1}+\boldsymbol\epsilon^q)$;
%\State $\mathbf{U}_t=\hat{\mathbf{U}}_{t-1} +\sum_{q=1}^K\frac{exp(-\frac{1}{\lambda}\alpha_t^\top\mathbf{J}^q) \bold\epsilon^q}{\sum_{q=1}^K exp(-\frac{1}{\lambda}\alpha_t^\top\mathbf{J}^q)}$;
\State Calculate estimated optimal control $\mathbf{U}_t$ in~\eqref{eq:3m_input} using the costs of the $K$ trajectories,
$(\alpha_t^\top\mathbf{J}^1,\cdots,\alpha_t^\top\mathbf{J}^K)$.
%\Statex \hspace{1cm} A
\end{algorithmic}
\end{algorithm}

\subsection{Multi-objective Multi-horizon Model Predictive Path Integral Control (3M)}\label{sec:3M}

Applying MPPI control in Section~\ref{sec:MPPI} to the feasibility maximization problem in Section~\ref{sec:MMPC}, we proposed the multi-objective multi-horizon model predictive path integral control (3M) algorithm. The proposed 3M algorithm is summarized in Algorithm~\ref{alg:1}, and explained below.

Given the current state $x_t$, the desired weight vector $\alpha_d$ can be calculated by~\eqref{eq:alphad} (line 1).
Given the desired weight vector $\alpha_d$, the $\alpha$ in~\eqref{eq:mmpc} can be selected close to the desired value $\alpha_d$ as in~\eqref{eq:alphaup} (line 3), where the input $\hat{\mathbf{U}}_{t-1}$ is constructed from the previous input $\mathbf{U}_{t-1}$ (line 2).

We sample $K$ trajectories of noises as follows (line 4):
\begin{align} \label{eq:Ksamplenoise}
    \boldsymbol\epsilon^q \in \RR^{(N+\frac{N(N-1)}{2}m)n_u}, \quad
    \epsilon^q(i) \sim {\mathcal{N}}(0,\Sigma)
\end{align}
for $q=1,\cdots,K$ and $i=1,\cdots,N+\frac{N(N-1)}{2}m$, where $\epsilon^q(i) \in \RR^{n_u}$ is the $i^{th}$ noise vector of $\boldsymbol\epsilon^q$.

For each sampled noise trajectory, we construct noise disturbed input $\hat{\mathbf{U}}_{t-1}+\boldsymbol\epsilon^q$, and evaluate the corresponding cost as (line 6):
\begin{align*}
    \alpha_t \mathbf{J}^q = \alpha_t \mathbf{J}(\hat{\mathbf{X}}_{t-1}^q,\hat{\mathbf{U}}_{t-1}+\boldsymbol\epsilon^q) \quad \text{for $q=1,\cdots,K$},
\end{align*}
where $\hat{\mathbf{X}}_{t-1}^q$ is the simulated state trajectory with the noise disturbed input $\hat{\mathbf{U}}_{t-1}+\boldsymbol\epsilon^q$ on the system~\eqref{eq:sysmodel} (line 5). The optimal control input is estimated by the weight-average of the costs calculated from the $K$ trajectories, $(\alpha_t^\top\mathbf{J}^1,\cdots,\alpha_t^\top\mathbf{J}^K)$, as (line 7):
\begin{align}\label{eq:3m_input}
    \mathbf{U}_t=\hat{\mathbf{U}}_{t-1}
+\sum_{q=1}^K \boldsymbol w^q{\boldsymbol\epsilon}^q,
\end{align}
where
\begin{align*}
    \boldsymbol w^q=\frac{exp(-\frac{1}{\lambda}\alpha_t^\top\mathbf{J}^q)
}{\sum_{q=1}^K exp(-\frac{1}{\lambda}\alpha_t^\top\mathbf{J}^q)} \in \RR_{>0}, {\rm \ \ for \ } q=1,\cdots,K.
\end{align*}
If $\gamma=0$, then the 3M algorithm reduces to the MPPI control. This is because the input toward the primary mission $\mathbf{U}^0$ optimizes the primary cost function $\mathbf{J}^0$ only, and the inputs toward the alternative missions $\mathbf{U}^i$ do not affect the optimization problem. Furthermore, $\alpha=[1,0,\cdots,0]^\top$ is the unique optimal solution of the problem~\eqref{eq:alphaup} and is recursively feasible if the initial condition is $\alpha_0=[1,0,\cdots,0]^\top$.

\section{Illustrative Examples}\label{sec:sim}
Simulations have been conducted for the motion planning problem of the unmanned aerial vehicle (UAV) and unmanned ground vehicle (UGV) with and without obstacles. The control objective of the simulations is to move the vehicle from the initial position $[0,0]$ to the primary destination $[10,10]$ in the $2$-D plane. The UAV simulations represent an urban drone delivery system, where the drone flies $200$-$500$ ft above ground and delivers a package to the primary destination. The alternative destinations can be seen as an emergency landing spot such as a safe rooftop. The UGV simulations adopt a more realistic dynamic system, where alternative destinations represent a place that can accommodate car repair and fueling/charging services. We use GPUs for parallel computation and discuss control frequency.

\subsection{Simulations on UAV}
We consider a double integrator model to simulate the UAV control problem:
\begin{align*}
    x_{k+1}=
    \left[
    \begin{array}{cccc}
     1 & 0 & 0.1 & 0   \\
     0 & 1 &  0 & 0.1 \\
     0 & 0 &  1 & 0   \\
     0 & 0 &  0 & 1
    \end{array}
    \right] x_k +
    \left[
    \begin{array}{cc}
     0  & 0  \\
     0  & 0  \\
     0.1 & 0  \\
     0  & 0.1 \\
    \end{array}
    \right]u_k ,
\end{align*}
where $x_k \in \RR^4$ represents the horizontal coordinate, vertical coordinate, horizontal velocity, and vertical velocity. The input $u_k \in \RR^2$ consists of horizontal and vertical accelerations. The initial condition is $x_0 = [0,0,0,0]^\top$ and the primary destination is $p^0 = [10,10,0,0]^\top$. Euclidean distance has been used for the distance metric $d$ in~\eqref{eq:MissionCompletetion} and~\eqref{eq:alphad}. The cost function in~\eqref{eq:costSmallJ} is constructed as $L^i(x,u)=(x-p^i)^\top Q (x-p^i) + u^\top R u$, $F^i(x) = (x-p^i)^\top Q (x-p^i)$ with $Q=\mathbb{I}$ and $R=\mathbb{I}$, where $\mathbb{I}$ is the identity matrix with a proper dimension. We use the desired weight vector $\alpha_d$ in~\eqref{eq:alphad} with $\lambda_{\alpha} = 1$. The cost function $h$ in the optimization problem~\eqref{eq:alphaup} is chosen as a quadratic function $h(\alpha - \alpha_d(x_k))=(\alpha - \alpha_d(x_k))^\top (\alpha - \alpha_d(x_k))$. The MPPI control parameters are $\Sigma={\mathbb{I}}$ and $\lambda = 0.5$. The weight $\gamma$ in~\eqref{eq:alphad}, control horizon $N$, and the number of sample trajectories $K$ are chosen differently for each simulation, and those parameters are presented in each figure. We compare the backup plan safety constrained control with the MPPI control ($\gamma=0$).

\subsubsection{Obstacle-free environment}
Two simulations have been conducted with two different alternative destinations ($m=2$); for the first simulation, mission states are given by $p^1=[2,6,0,0]^\top$ and $p^2 = [8,6,0,0]^\top$; for the second simulation, mission states are $p^1=[2,8,0,0]^\top$ and $p^2 = [6,12,0,0]^\top$.

\begin{figure}[t]
    \centering \vspace{0.2cm}
    \includegraphics[width=0.85\linewidth]{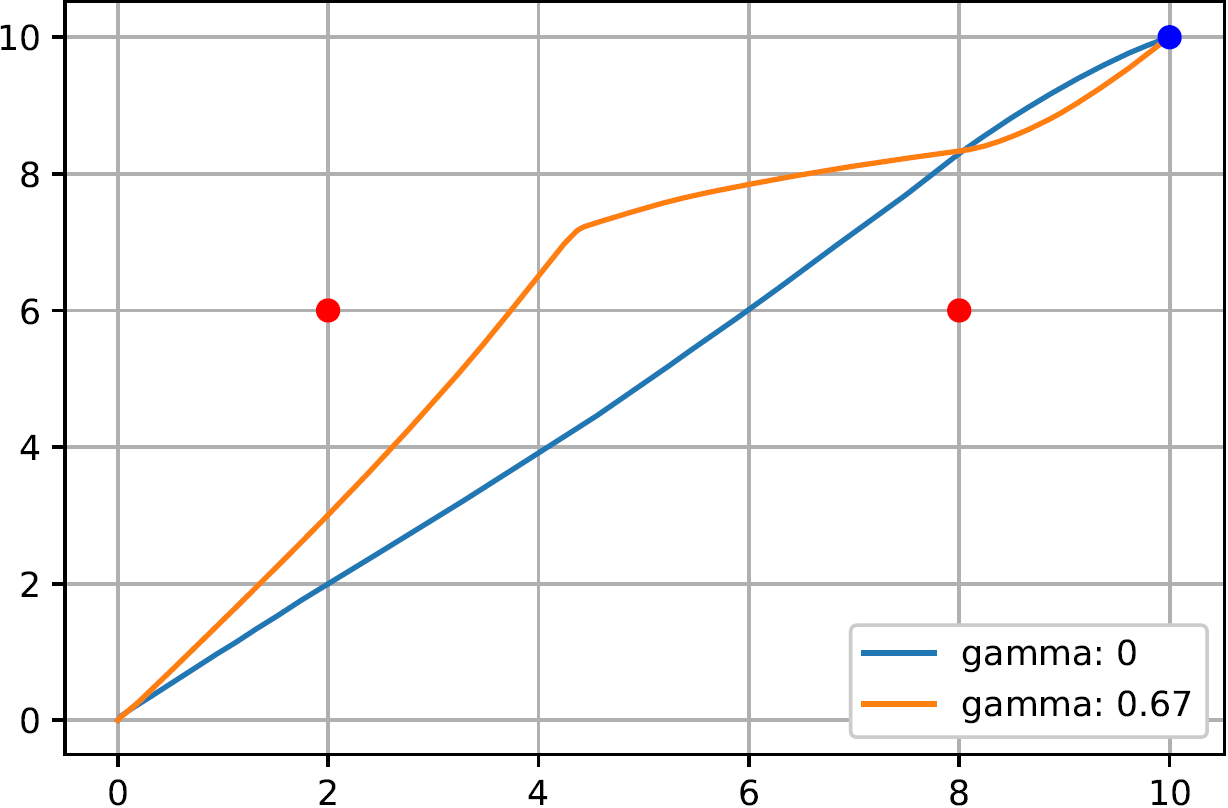}
    \caption{UAV simulation in an obstacle-free environment with $N = 10$ and $K=1000$. The orange line is the executed state trajectory of the UAV for backup plan constrained control, and the blue line is the trajectory for the regular MPPI toward the primary destination.}\vspace{-0.3cm}
    \label{fig:sim_UAV}
\end{figure}

\begin{figure}[t]
    \centering
    \includegraphics[width=0.85\linewidth]{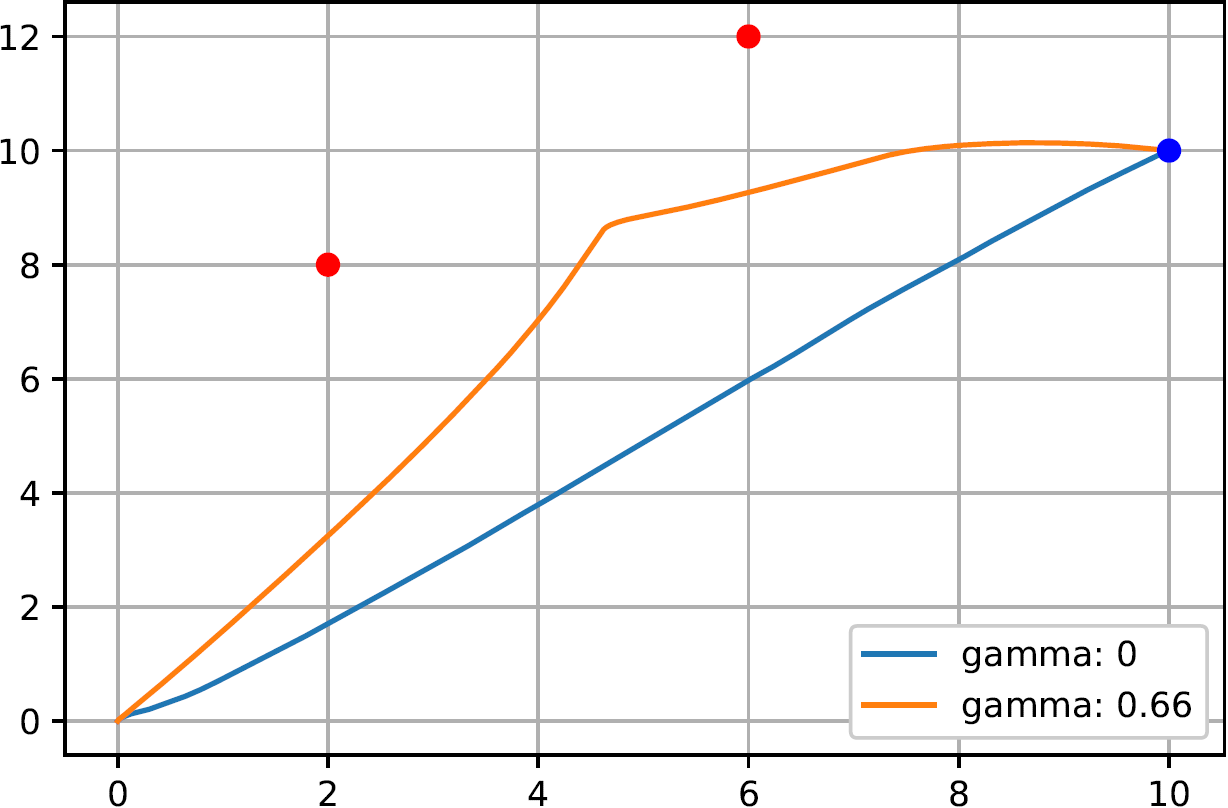}
    \caption{UAV simulation in an obstacle-free environment with $N = 10$ and $K=1000$.}\vspace{-0.6cm}
    \label{fig:sim_UAV2}
\end{figure}

The results are presented in Figures~\ref{fig:sim_UAV} and~\ref{fig:sim_UAV2}, where the primary destination is marked in a blue dot, and alternative destinations are marked in red dots. The UAV with MPPI control ($\gamma=0$) flies to the primary destination directly as expected. The UAV with the 3M algorithm makes a detour to the primary destination in both cases, flying near alternative destinations. The detour trajectory is safer in the backup plan sense, providing a shorter path toward one of the alternative destinations when an emergency landing is in need.

\begin{figure}[t]
    \centering \vspace{0.2cm}
    \includegraphics[width=0.85\linewidth]{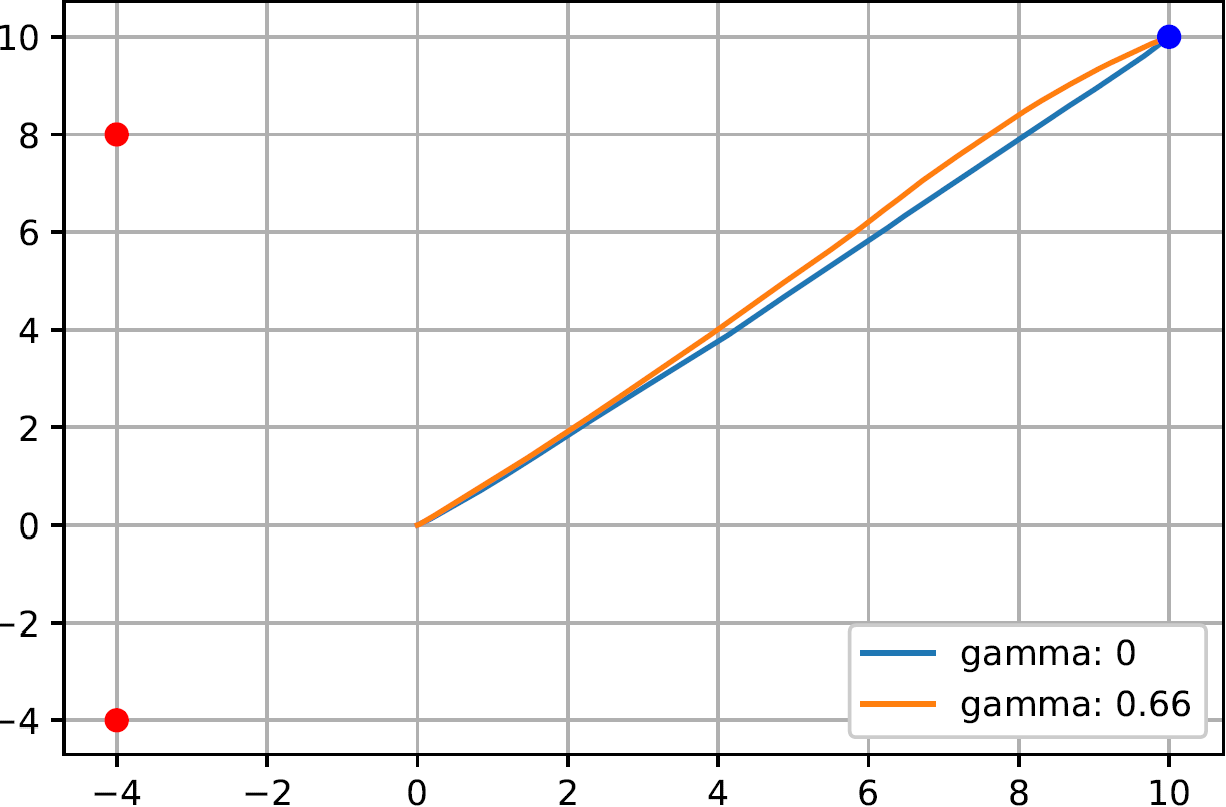}
    \caption{UAV simulation in an obstacle-free environment with $N = 10$ and $K=1000$. The alternative destinations are located in the opposite direction to the primary destination.}
    \label{fig:sim_UAVop}\vspace{-0.3cm}
\end{figure}

It is essential to choose a set of suitable alternative missions to be partially aligned with the primary mission. If not, the proposed 3M algorithm automatically less considers conflicting alternative missions over time by adopting a distance-based update law for the desired weight vector. For example, if the alternative destinations are located in the opposite direction to the primary destination ($p^1=[-4,-4,0,0]^\top$ and $p^2=[-4,8,0,0]^\top$), the trajectory with $\gamma=0.66$ has a subtle difference from that with $\gamma=0$ as shown in Figure~\ref{fig:sim_UAVop}.


\subsubsection{Environment with obstacle}
Two alternative destinations are located at $p^1 = [2,6,0,0]^\top$ and $p^2 = [6,12,0,0]^\top$. A soft constraint renders the collision avoidance constraint. 

\begin{figure}[t]
    \centering
    \includegraphics[width=0.85\linewidth]{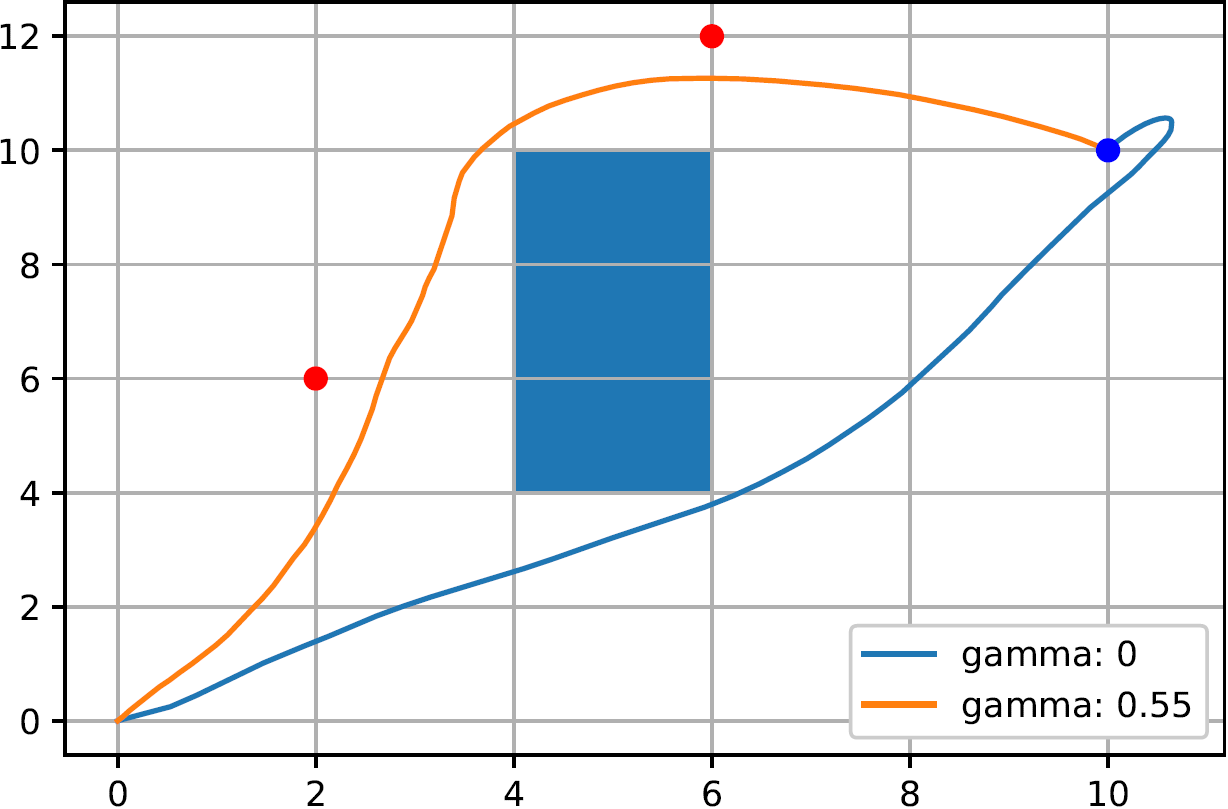}
    \caption{UAV simulation in an environment with obstacles with $N = 20$ and $K=1000$.}\vspace{-0.6cm}
    \label{fig:sim_UAVobstacle}
\end{figure}

Figure~\ref{fig:sim_UAVobstacle} shows the simulation results, where the blue boxes are obstacles. The UAV flies near the safe rooftops, avoiding obstacles instead of choosing the shortest path. When $\gamma=0$, the UAV arrives at the destination with a hook shape trajectory to land with zero velocity.

\subsection{Simulations on UGV}
The UGV simulations use the simple car dynamic model described by (Chapter 13.1.2 in~\cite{lavalle2006planning}):
\begin{align*}
    p^x_{k+1} &= p^x_{k} + (v_t \cos \theta_k)\delta \\
    p^y_{k+1} &= p^y_{k} + (v_t \sin \theta_k)\delta \\
    \theta_{k+1} &= \theta_{k} + (\frac{v_t}{L} \tan \phi_k)\delta,
\end{align*}
where $x_k = [p^x_k,p^y_k,\theta_k]^\top \in \RR^3$ represents horizontal coordinate, vertical coordinate, and heading angle, respectively. Input $u_k=[v_k,\phi_k]^\top$ consists of velocity and steering angle. Parameter $L=0.2$ is a wheelbase, and $\delta=0.1$ is the time step. Other functions and parameters remain unchanged, if not specified. There are two alternative destinations located at $p^1 = [2,6,0,0]^\top$ and $p^2 = [6,12,0,0]^\top$.

\begin{figure}[t]
    \centering \vspace{0.2cm}
    \includegraphics[width=0.85\linewidth]{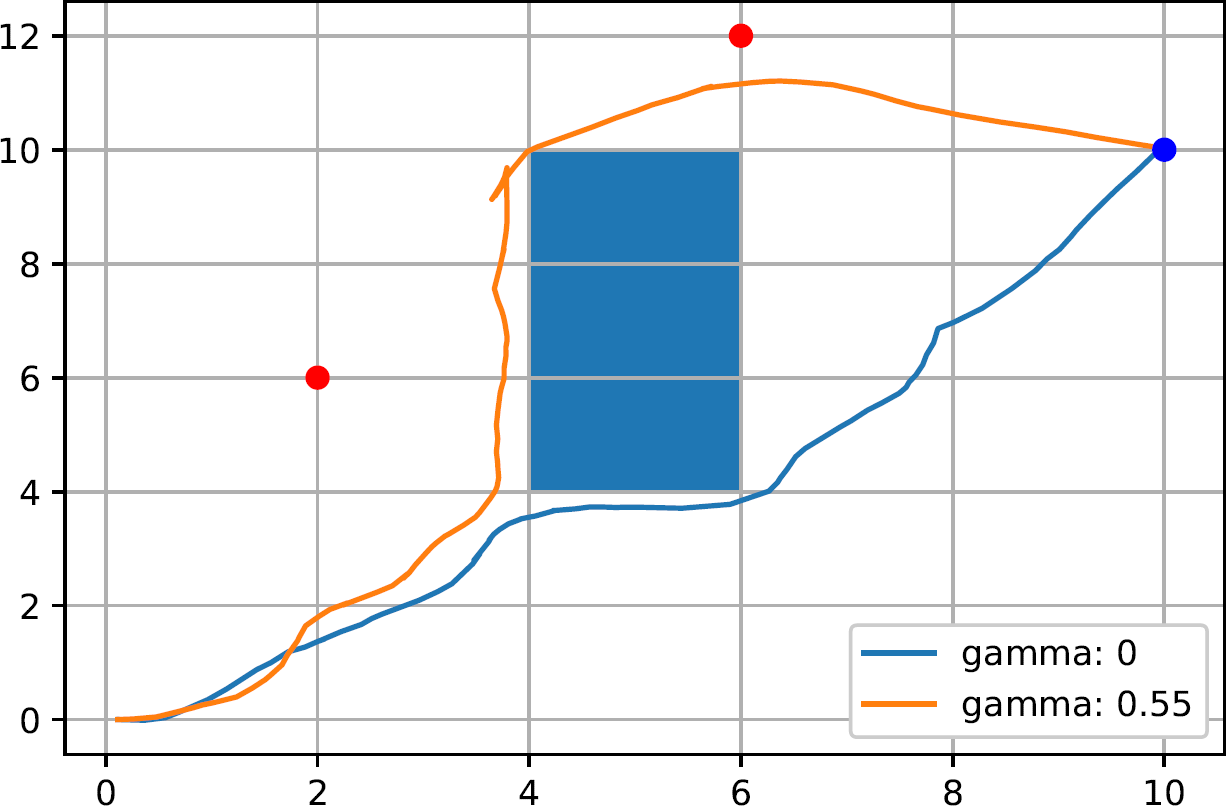}
    \caption{UGV simulation in an environment with obstacles with $N = 10$ and $K=10000$.}\vspace{-0.3cm}
    \label{fig:sim_UGVobstacle}
\end{figure}

The simulation result in Figure~\ref{fig:sim_UGVobstacle} shows that the executed trajectory remains similar even when the dynamic system model has been changed. That is, the system tries to enhance the backup plan safety. The path is not smooth compared to those of the double integrator model. This is because the simple car model is a nonlinear model with non-holonomic constraints that restrict the motion, e.g., it cannot make a turn when $v_t=0$, and cannot move to the lateral direction.
\begin{figure}[t]
    \centering 
    \includegraphics[width=0.85\linewidth]{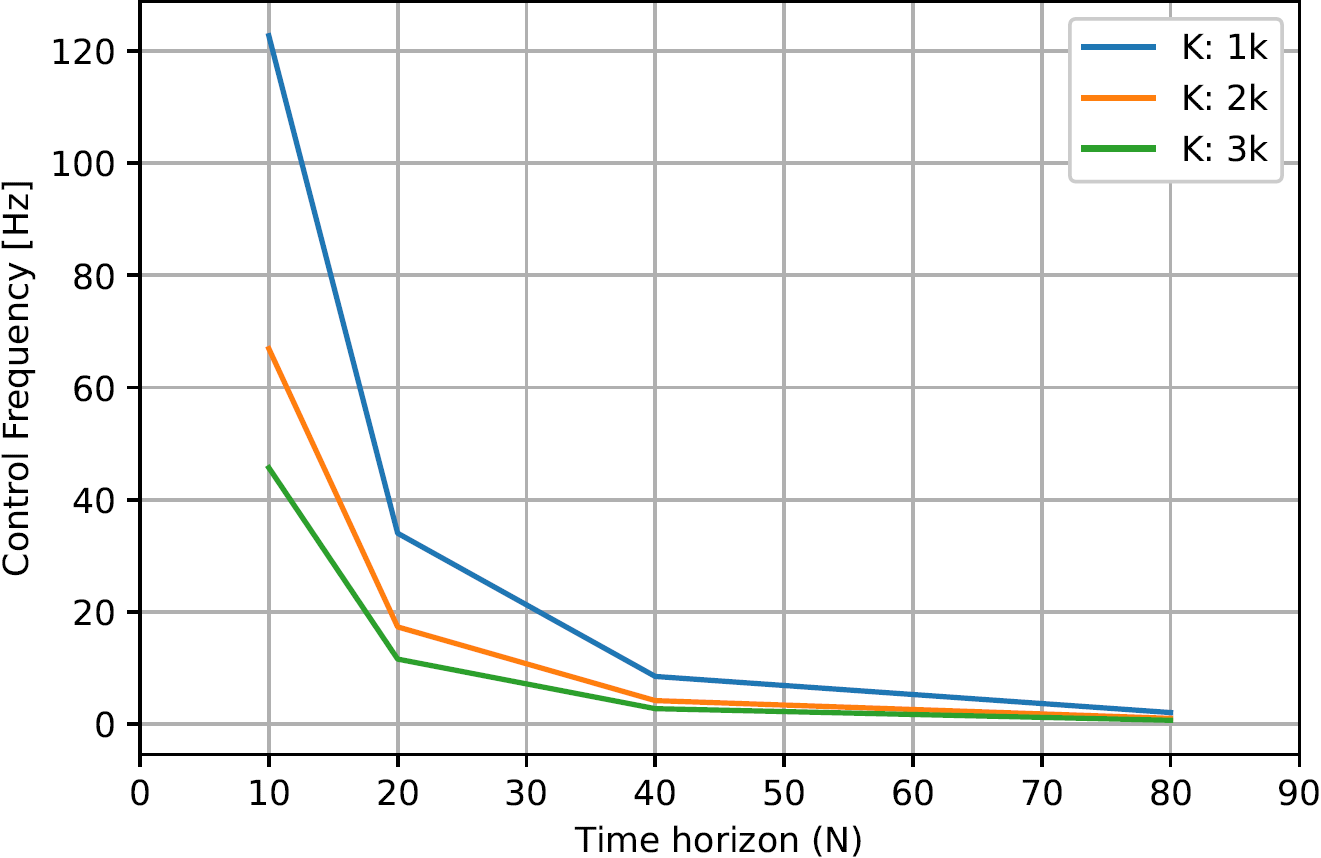}
    \caption{Control frequency with respect to the prediction horizon $N$.}\vspace{-0.6cm}
    \label{fig:sim_A2}
\end{figure}

\subsection{Analysis and Discussion}

This section presents simulation results regarding computation, cost, and standard deviation.

Figure~\ref{fig:sim_A2} shows control frequency with respect to the prediction horizon $N$. We have used a desktop computer with \emph{AMD Ryzen 5 3600} CPU, 16GB RAM, and \emph{NVIDIA GeForce RTX 2070} GPU for running the simulation with the proposed 3M algorithm. As the number of inputs increases in the order of $N^2$, the computational complexity for the MPPI part is expected to increase in $N^2$ as well. Using GPUs, it is real-time implementable up to $N=40$.

\begin{figure}[t]
    \centering \vspace{0.2cm}
    \includegraphics[width=0.85\linewidth]{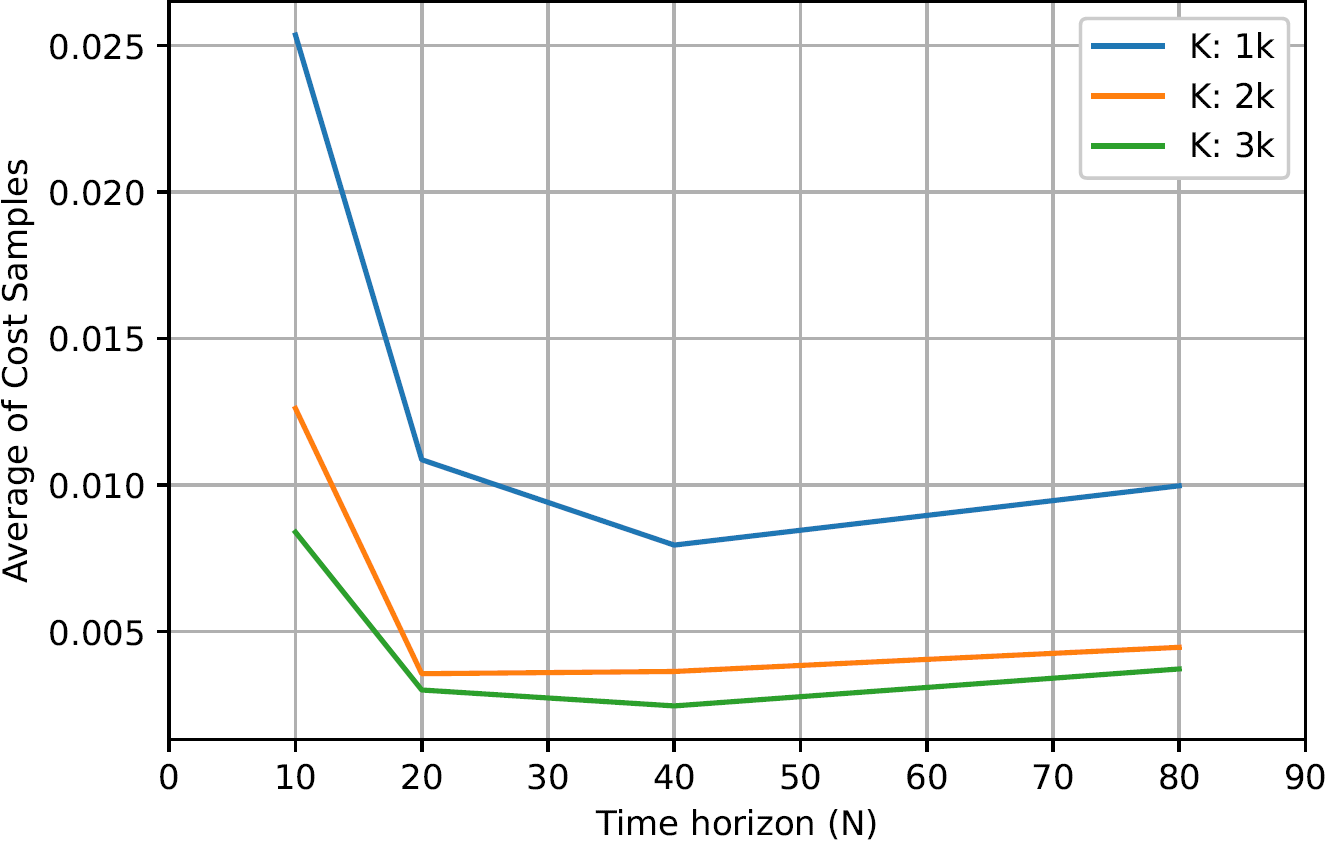}
    \caption{Average of the cost samples $\alpha^\top \mathbf{J}$.}\vspace{-0.3cm}
    \label{fig:sim_A3}
\end{figure}

Figure~\ref{fig:sim_A3} shows that the average cost remains high when it is near-sighted (i.e., $N$ is small). The cost decreases as $N$ increases until $N=40$. After then, the cost starts to increase because $\mathbf{J}^i$ is the sum of $N$ horizon costs.

\begin{figure}[t]
    \centering
    \includegraphics[width=0.85\linewidth]{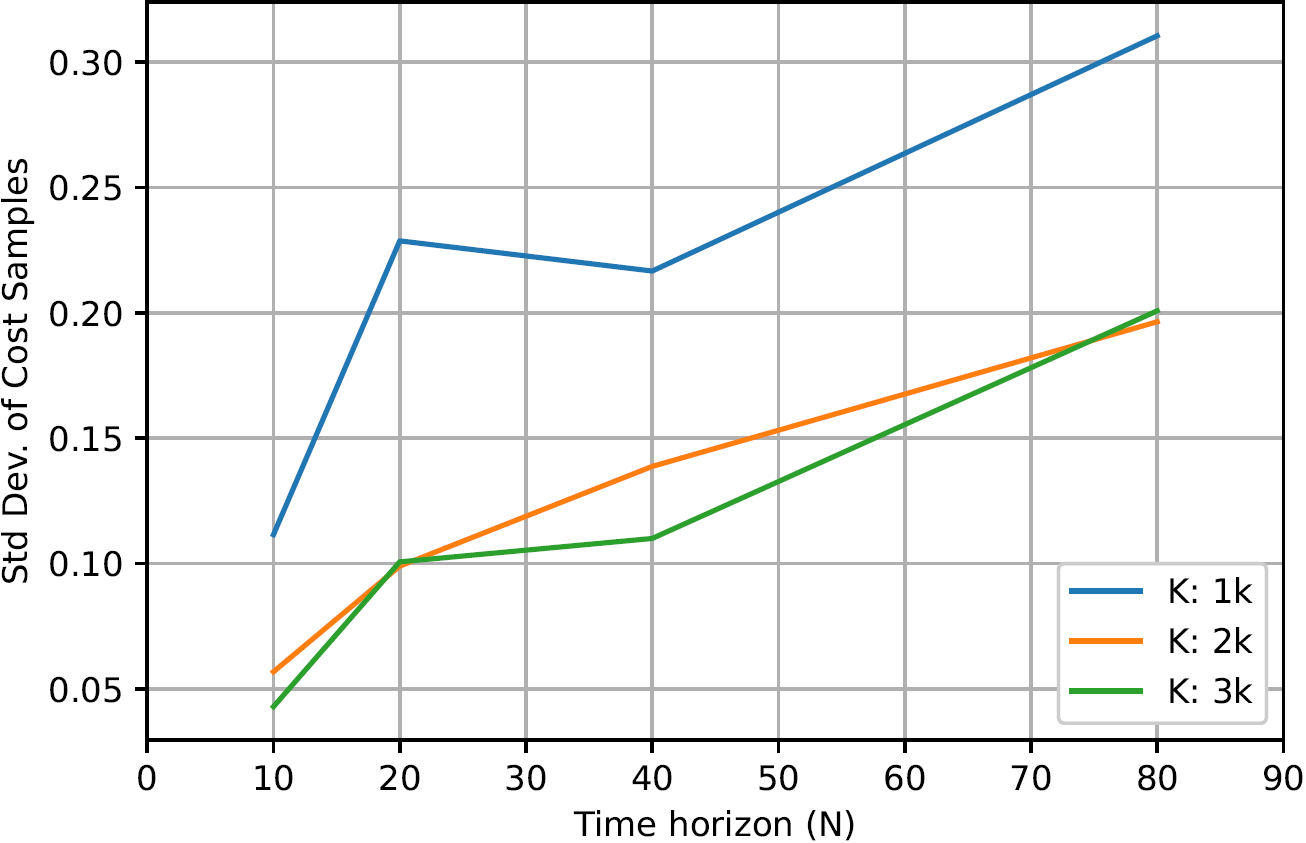}
    \caption{Standard deviation of the cost samples $\alpha^\top \mathbf{J}$.}\vspace{-0.62cm}
    \label{fig:sim_A4}
\end{figure}

The standard deviation of the cost distribution decreases as $K$ increases as shown in Figure~\ref{fig:sim_A4}. This result is consistent with the known result in statistics that the variance of importance sampling (IS)\footnote{The MPPI uses normal distribution samples to estimate the distribution of the cost of optimized control. This method of using a different distribution for estimating target distribution is called importance sampling in statistics.} estimator depends on the number of samples, i.e., $\sigma_\text{IS est.} = \sigma_q/K$, where $\sigma_q$ is the standard deviation of the random sampling (equation (6.5) in \cite{owen2013monte}). Also, the increasing variance in $N$ can be explained by an accumulation of variance due to the addition of random variables.

\section{Conclusion}
The motivation of this work is to enable the development of the new safety concept for autonomous systems, while the current one has been limited to collision avoidance. For instance, many industries utilize the constrained motion planning on their systems and could benefit from collision-free for safety, such as indoor navigation robots, follow-filming drones, and self-driving cars. But more often than not, only considering the collision avoidance at the planning level is not enough for safety-critical systems since the primary mission may not be feasible under some unforeseen conditions. This work addresses these concerns by introducing a novel safety concept: backup plan safety that also considers the feasibility of the alternative missions. Otherwise, searching for alternative missions after finding the primary mission is not feasible could lead to dangerous consequences. 

This paper studies a novel safety concept, backup plan safety. To fulfill the safety in the control problem, we formulate the control problem as a feasibility maximization problem, which is addressed by multi-horizon multi-objective model predictive path integral control, which adopts additional control horizons toward the alternative missions on top of the control horizon toward the primary mission. Simulations of  aerial vehicle and ground vehicle control problems illustrate the new concept of backup plan safety and the performance of the proposed algorithms.

\bibliographystyle{ieeetr}
\bibliography{a_reference}

\end{document}

%% file: Misc/a_macros.tex
\usepackage{cite}

\usepackage[framemethod=tikz]{mdframed}

\usepackage{hyperref}
\hypersetup{
	colorlinks=true,
	linktoc=all,
	linkcolor=blue, %blue,
	citecolor=blue
}

\usepackage{amsmath,amssymb,amsfonts}
\usepackage{algorithm,algorithmicx,algpseudocode}
\usepackage{graphicx}
\usepackage{subcaption}
\usepackage{mwe}
\usepackage{textcomp}
\usepackage{xcolor,soul}
\definecolor{Blue}{RGB}{88, 105, 225}
\definecolor{QOrange}{RGB}{255, 153, 51}
\definecolor{Orange}{RGB}{255, 153, 50}

\usepackage{dsfont}
\usepackage{bbm}

\DeclareMathOperator{\argmin}{arg\,min}

\newtheorem{definition}{\bf Definition}[section]
\newtheorem{remark}{\bf Remark}[section]

\newcommand{\nnum}{\nonumber}

\newcommand{\RR}{\mathbb{R}}

\algnewcommand{\LineComment}[1]{\Statex \(\triangleright\) #1}

\newcommand\oprocendsymbol{\hbox{$\blacksquare$}}
\newcommand\oprocend{\relax\ifmmode\else\unskip\hfill\fi\oprocendsymbol}

\DeclareMathAlphabet{\mathsfit}{\encodingdefault}{\sfdefault}{m}{sl}
\SetMathAlphabet{\mathsfit}{bold}{\encodingdefault}{\sfdefault}{bx}{n}